\documentclass[12pt,preprint]{aastex}

\shorttitle{Spitzer Observations of Cool WDs}
\shortauthors{Kilic et al.}

\begin{document}

\title{The Mystery Deepens: Spitzer Observations of Cool White Dwarfs}

\author{Mukremin Kilic\altaffilmark{1}, Ted von Hippel\altaffilmark{1}, Fergal Mullally\altaffilmark{1}, William T. Reach\altaffilmark{2}, Marc J. Kuchner\altaffilmark{3}, D. E. Winget \altaffilmark{1}, and Adam Burrows\altaffilmark{4}}

\altaffiltext{1}{Department of Astronomy, University of Texas at Austin, 1 University Station
C1400, Austin, TX 78712; kilic@astro.as.utexas.edu}
\altaffiltext{2}{Spitzer Science Center, MS 220-6, California Institute of Technology, Pasadena, CA 91125}
\altaffiltext{3}{NASA Goddard Space Flight Center, Greenbelt, MD 20771}
\altaffiltext{4}{Department of Astronomy and Steward Observatory, University of Arizona, 933 North Cherry Avenue, Tucson, AZ 85721}

\begin{abstract}

We present 4.5$\mu$m and 8$\mu$m photometric observations of 18 cool white dwarfs obtained with the
$Spitzer$ Space Telescope. Our observations demonstrate that four white dwarfs with $T_{\rm eff}<$ 6000 K show slightly
depressed mid-infrared fluxes relative to white dwarf models. In addition, another white dwarf with a peculiar optical
and near-infrared spectral energy distribution (LHS 1126) is found to display significant flux deficits in Spitzer
observations. These mid-infrared flux deficits are not predicted by the current white dwarf models including collision
induced absorption due to molecular hydrogen. We postulate that either the collision induced absorption calculations
are incomplete or there are other unrecognized physical processes occuring in cool white dwarf atmospheres. The spectral energy
distribution of LHS 1126 surprisingly fits a Rayleigh-Jeans spectrum in the infrared, mimicking a hot white dwarf
with effective temperature well in excess of 10$^5$ K. This implies that the source of this flux deficit is probably
not molecular absorption but some other process.

\end{abstract}


\section{Introduction}

The Sloan Digital Sky Survey (Adelman-McCarthy et al. 2005) has increased the number of known field cool white dwarfs from 
tens of objects to thousands (Kilic et al. 2006; Harris et al. 2006). In addition, Hubble Space
Telescope observations of the globular clusters M4 (Hansen et al. 2004) and $\omega$ Cen (Monelli et al. 2005) resulted in the
discovery of more than two thousand white dwarfs. A careful analysis of these large datasets require
a complete understanding of the structure and evolution of the white dwarf stars. The main uncertainty in the white dwarf
luminosity functions derived from these datasets is caused by the inadequate description of energy transport in white dwarf atmospheres,
which affects both the cooling rate and observational appearence of these stars (Hansen 1998).

Cool white dwarfs have atmospheres dominated by hydrogen or helium. Both hydrogen and helium are neutral below 5000 K and the
primary opacity source in H-rich cool ($T_{\rm eff}\leq5500$ K) white dwarf atmospheres is believed to be collision induced absorption (hereafter CIA; Frommhold 1993)
of molecular hydrogen (Bergeron et al. 1995; Hansen 1998; Saumon \& Jacobson 1999). H-rich white dwarfs are predicted to become redder as they
cool until the effects of CIA become significant below 5500 K. CIA opacity is strongly wavelength dependent and is expected
to produce broad absorption features in the near-infrared.
This flux deficiency can be seen in several stars, designated as ultracool white dwarfs (Oppenheimer et al. 2001; Harris et al. 2001; Gates et al. 2004; Farihi 2004; Farihi 2005). 

There are two known opacity mechanisms in pure He-rich white dwarf atmospheres: Rayleigh scattering and He$^-$ free-free absorption.
Rayleigh scattering is thought to be the dominant opacity source and therefore He-rich white dwarfs are expected to have
blackbody like spectral energy distributions. On the other hand, Kowalski, Saumon \& Mazevet (2005) argued that He$^-$ free-free
absorption may be 2-3 orders of magnitude more significant, which would make it the dominant opacity source.
In addition, Kowalski \& Saumon (2004) showed that the atmosphere of He-rich white dwarfs with $T_{\rm eff}<$ 10000 K should
be treated as a dense fluid rather than an ideal gas, and that refraction effects become important. The density can be as
high as 2 g cm$^{-3}$ in these atmospheres and the index of refraction departs significantly from unity.
Kowalski, Saumon, \& Mazevet (2005) used their updated models to match the observed sequence of cool white dwarfs from
Bergeron, Ruiz, \& Legget (1997; hereafter BRL) and suggested that the coolest white dwarfs have mixed H/He atmospheres. 

Bergeron \& Leggett (2002) tried to fit the optical and near-infrared photometry of two ultracool white dwarfs, LHS 3250 and SDSS 1337+00,
and ruled out their models for pure hydrogen atmospheric composition for these stars. 
They found that the overall spectral energy distributions (SEDs)
of these stars can be better fitted with mixed H/He models, yet the peak of the SEDs near 6000 \AA\ is predicted to be too
narrow. There are only three white dwarfs with significant CIA that fits the current white dwarf models in the
optical and near-infrared. WD0346+246 (Oppenheimer et al. 2001) and GD392B (Farihi 2004) both require low gravity (log $g\sim$7),
mixed H/He model atmosphere solutions. In addition, Bergeron et al. (1994) found a 5400 K, $log$ g=7.9, and $log$
N(He)/N(H) = 0.8 model atmosphere solution for LHS1126. BRL used new CIA opacity calculations and suggested that the helium
abundance in this white dwarf is $log$ N(He)/N(H) = 1.86.

The $Spitzer$ Space Telescope (Werner et al. 2004) opened a new window into the Universe by enabling accurate mid-infrared
photometry of faint objects (microjansky-level sensitivity). In order to understand the CIA opacity, and other unrecognized
sources of opacity in
cool white dwarf atmospheres, we used the $Spitzer$ Space Telescope to observe nearby, relatively bright, cool white dwarfs.
In this paper, we present our mid-infrared photometry for 18 cool white dwarfs including LHS 1126. Our observations
are discussed in \S 2, while an analysis of these data and results are discussed in \S 3. 

\section{Target Selection and Observations}

We selected our targets from the spectroscopically-confirmed
white dwarf catalog (McCook \& Sion 1999) and required that they have $K_S\leq$ 15 mag. as measured by the Two Micron All Sky Survey (2MASS; Cutri et al. 2003).
Observations reported here were obtained as part of the Spitzer Cycle 1 GO-Program 2313 (PI: Marc Kuchner). We used the
Infrared Array Camera (IRAC; Fazio et al. 2004) to obtain 4.5 and 8$\mu$m (effective wavelengths of 4.493 and 7.782 $\mu$m)
photometry of 18 cool white dwarfs with
5000 K $<T_{\rm eff}<$ 9000 K. An integration time of 30 seconds per dither, with five dithers for each target, was used
(150 seconds total integration time).

We used the products of the Spitzer Science Center pipeline, the Basic Calibrated Data (BCD) Frames and the Post-BCD frames
(mosaics), for our analysis. We used the IRAF\footnote{IRAF is distributed by the National Optical Astronomy Observatory, which is operated by the Association of Universities for Research in Astronomy (AURA), Inc., under cooperative agreement with the National Science Foundation.}
PHOT routine to perform aperture photometry on individual BCD frames. Experience showed that the Point Response Function
for the IRAC instrument is not well defined, and we obtained better results with aperture photometry than with PSF-fitting photometry.
In order to maximize the signal-to-noise ratio, we used a 5 pixel aperture for bright, isolated objects, and 2 or 3 pixel
apertures for faint objects or objects in crowded fields. We corrected the resultant fluxes by the aperture correction factors
determined by the IRAC team (see the IRAC Data Handbook).
We checked the results from 2, 3, and 5 pixel apertures for each object, and found them to be consistent within the errors.

Following the IRAC calibration procedure, corrections for the location of the source in the array
were taken into account before averaging the fluxes of each of the five frames at each wavelength.
We also performed photometry on the mosaic images and found the results to be consistent with the photometry from individual frames.
We divided the estimated fluxes by the color corrections for a Rayleigh-Jeans
spectrum (Reach et al. 2005a). These corrections are 1.0121 and 1.0339 for the 4.5 $\mu$m and 8$\mu$m-bands, respectively.
Based on the work of Reach et al. (2005a), we expect that our IRAC photometry is calibrated to an accuracy of 3\%. The
average fluxes measured from the Spitzer images along
with the 2MASS photometry, spectral types, and temperatures for our objects are given in Table 1.
The error bars were estimated from the observed scatter in the 5 images (corresponding to 5 dither positions) plus the 3\%
absolute calibration error, added in quadrature.

\section{Results}

Figure 1 shows the optical and infrared spectral energy distributions (filled triangles) of the 18 cool
white dwarfs (ordered in $T_{\rm eff}$) that we observed with the IRAC instrument. Most of our objects were observed by Bergeron, Leggett, \& Ruiz (2001;
hereafter BLR), and therefore have accurate $BVRI$ photometry, and temperature determinations. For the four DA white dwarfs not included
in BLR's analysis (WD0018-267, WD0141-675, WD0839-327, and WD1223-659), we used the $UBV$ photometry from McCook \& Sion (1999)
catalog. The near-infrared photometry comes from the 2MASS All-Sky Point Source Catalog.

The expected fluxes from synthetic photometry of white dwarf model atmospheres (integrated over the filter
bandpasses; kindly made available to us by D. Saumon and
D. Koester) are shown as open circles in Figure 1. These models include CIA opacities for H-rich objects with $T_{\rm eff}<$ 7000 K.
We normalized the model atmospheres to the observed SEDs in the V-band (0.55 $\mu$m), with the exception of WD1748+708 (G240-72).
WD1748+708 shows a 15\% deep, 2000 \AA\ wide absorption feature centered at $\sim$ 5300 \AA\ that affects its $BVR$ photometry (BRL).
Hence, we matched its model to the observations in the $I$-band. Solid lines represent the same models normalized to the observed
SEDs in the $J$-band. Changing the normalization of the models to the $J$-band does
not change our results significantly since the models and observations agree fairly well between the $V$ and $J$ bands.

In order to derive new $T_{\rm eff}$ values for our objects, we used pure-H and pure-He white dwarf models (log $g=8$)
and employed a $\chi^2$ minimization technique to fit $V$, $RI$ (if available), $JHK$, 4.5$\mu$m, and 8$\mu$m photometry.
$V$ and $R$ photometry was omitted in our analysis of WD1748+708 because of the strong absorption feature observed in these filters.
Our estimated $T_{\rm eff}$ values for the 8 DA white
dwarfs that are in common with the BLR sample are on average slightly hotter (+66 $\pm$ 81 K)
than the BLR temperatures, though they are consistent within the errors.

\subsection{White Dwarfs with $T_{\rm eff}<$ 7000 K}

A comparison of observed versus expected SEDs from white dwarf model atmospheres shows that 
all of the cool white dwarfs with $T_{\rm eff}<$ 6000 K (except the DZ white dwarf WD0552-041) show
slightly depressed mid-infrared fluxes relative to white dwarf models. In addition,
WD0038-226 (LHS 1126) displays significant flux deficits (more than 5$\sigma$ in each band) in the mid-infrared.
There are also two more stars with 6000 K $<T_{\rm eff}<$ 7000 K (WD0009+501 and WD0141-675) that display $\sim1\sigma$ flux deficits.
Figures 2 and 3 show the observed flux ratios in different filters for the white dwarfs in our sample. H and He rich white dwarfs
are shown as filled and open circles, respectively. White dwarfs with uncertain spectral types (LHS 1126 and WD1748+708) are shown
as filled triangles. Cooling tracks for H (solid line) and He rich (dashed line) white dwarfs,
along with a blackbody cooling track (dotted line) are also shown. Figure 2 shows that H-rich white dwarfs with $T_{\rm eff}<7000$ K
emit less flux at 8$\mu$m than predicted from the white dwarf models. In addition, their infrared colors (Figure 3)
require most of them to be warmer than 6500 K; they mimic warmer/bluer objects in the infrared.

LHS 1126 shows strong molecular features in the optical that are thought to be due to C$_2$H. These absorption features are
blueshifted by about 150 \AA\ compared to the C$_2$ swan bands. Several investigators tried to explain these features in terms
of either pressure shifts in a He-dominated atmosphere or magnetic displacements of the swan bands (see Schmidt et al. 1995).
However, Bergeron et al. (1994) and Schmidt et al. (1995) showed that both scenarios failed for this star. In addition, Schmidt
et al. (1995) suggested that C$_2$H is the most probable molecule to form under the conditions in LHS 1126.
WD1748+708 shows a very broad feature in the optical (see the discussion above) and has a $\sim$200 MG magnetic field
(Angel 1977). BLR suggested that this broad feature maybe explained as C$_2$H molecular feature broadened by the strong magnetic field.
Hence, all of the stars with mid-infrared flux deficits have either H-rich atmospheres or show trace amounts of hydrogen.

The coolest white dwarf in our sample is a DZ white dwarf, WD0552-041, with an estimated $T_{\rm eff}=$ 5016 K. The observed
SED fits the model predictions fairly well for this star with the exception of $B$ photometry which is probably affected by the
metals in the photosphere (Wolff et al. 2002). The origin of metals in white dwarf atmospheres has been a mystery
for a long time (Zuckerman et al. 2003). The discovery of debris disks around the white dwarfs G29-38
(Zuckerman \& Becklin 1987; Reach et al. 2005b) and GD362 (Becklin et al. 2005; Kilic et al. 2005) suggest that accretion
from a debris disk may explain the observed metal abundances in DAZ white dwarfs. Nevertheless, we do not see any mid-infrared excess
around the cool DZ white dwarf WD0552-041.

\subsection{White Dwarfs with $T_{\rm eff}>$ 7000 K}

The remaining ten objects with $T_{\rm eff}>$ 7000 K do not show any excess or deficit in their flux distributions.
There are three known or suspected double degenerate stars in our sample (WD0101+048, WD0126+101, and WD0839-327;
Maxted et al. 2000). We do not see any evidence of a composite SED for these objects;
if they are equal mass binaries, the primary and secondary stars must have similar temperatures.
Limits on possible substellar companions or debris disks around these stars as well as $\sim$100 other white dwarfs will be
discussed in a future publication (von Hippel et al., in preparation).
One caveat seen in Figure 1 is that WD1121+216 is near a brighter
source, and even a 2 pixel aperture on the white dwarf is contaminated by the light from the nearby star. Hence, we
provide only an upper limit for 8$\mu$m photometry of this object.

WD2140+207 is a DQ star showing molecular C$_2$ swan bands in the optical. The presence of carbon in cool He white dwarf atmospheres
is thought to be the result of convective dredge up (Pelletier et al. 1986).
Figure 1 shows that the spectral energy distribution of WD2140+207 fits a blackbody distribution reasonably well. The absence of metal
lines in the optical spectrum of this star, plus the absence of excess infrared radiation provide further evidence
that the carbon was not accreted from the interstellar medium or a circumstellar debris disk. 
One caveat between our analysis and BLR's analysis is that using pure-He model atmospheres, BLR estimated $T_{\rm eff}=$ 8860
$\pm$ 300 K for this star. Fitting $VRIJHK$, 4.5$\mu$m and 8$\mu$m photometry of this star with a blackbody, we estimate
the temperature of this star to be 9407 K. None of the other stars in our sample that are in common with the BLR sample show
this discrepancy. Dufour et al. (2005) reanalyzed the DQ stars in the BLR sample and demonstrated that the inclusion of carbon
in model atmosphere calculations reduces the estimated effective temperatures and surface gravities for DQ white dwarfs.
They obtained even a lower $T_{\rm eff}$ measurement (8200 $\pm$ 250 K) for WD2140+207.

\section{Discussion}

Our Spitzer observations showed that all H-rich white dwarfs with $T_{\rm eff}<7000$ K show slight mid-infrared flux deficits.
Having several stars with small deficits makes these deficits significant. Moreover, LHS 1126 shows significantly depressed
mid-infrared fluxes relative to white dwarf models. Debes et al. (2005, private communication) have also found 10-20\% flux deficits
($\geq3\sigma$ significance) at 4.5, 5.6, and 8$\mu$m in two DAZ white dwarfs with $T_{\rm eff}=$ 6820 K and 7310 K.
Combining the facts that WD0552-041 (with a pure-He atmosphere) does not show a clear flux deficit, and all H-rich white dwarfs cooler
than 7000 K do exhibit mid-infrared flux deficits, LHS1126 and WD1748+708 are likely to have mixed H/He atmospheres. This is also
consistent with Bergeron et al.'s (1994) and Wolff et al.'s (2002) analysis. However, one question remains to be answered:
can these flux deficits be explained by CIA?

The atmospheres of M, L, and T dwarfs are also expected to exhibit CIA (Borysow et al. 1997).
Roellig et al. (2004) obtained mid infrared spectroscopy of M, L, and T dwarfs in the 5 -- 15 $\mu$m range. Their Figure 2 shows
that the observed spectra are in good agreement with the model atmopshere calculations, with only a few minor deviations.
Hence, CIA calculations for low density atmospheres seem to be accurate.

The near-infrared (1--2 $\mu$m) flux deficit in LHS 1126 was discovered earlier by Wickramasinghe et al. (1982). Bergeron et al. (1994)
and BRL explained this deficit as CIA by molecular hydrogen due to collisions with helium and found the H/He
ratio to be $\sim$0.01. Wolff et al. (2002) used Faint Object Spectrograph data plus optical
and infrared photometry of LHS1126 to model this star's SED in the 0.2 -- 2.2 $\mu$m range. They found that the hydrogen
abundance reported by Bergeron et al. (1994) would result in an extremely strong Lyman $\alpha$ absorption, and the SED is best fitted
with an abundance ratio of H/He = 3 $\times$ 10$^{-6}$. Wolff et al. (2002) and Bergeron et al. (1994) models do not give a
consistent picture for the H/He ratio and neither model is adequate.
We revisit this problem by extending our wavelength coverage to 8$\mu$m.
Figure 4 shows the ultraviolet spectrum (kindly made available to us by D. Koester) and optical and infrared photometry of
LHS 1126 along with a 5400 K blackbody (dotted line). Mixed atmosphere white dwarf models (kindly made available to us by D. Saumon)
with $T_{\rm eff}=5400$ K, $log$ g= 7.9, and $log$ N(He)/N(H) = 1.5 (lower solid line) and $log$ N(He)/N(H) = 1 (upper solid line)
are also shown.
These white dwarf models include CIA opacities but not Lyman $\alpha$ absorption, therefore they cannot be used to match
the ultraviolet data. If we just use the optical and near-infrared photometry (as in Bergeron et al. 1994), we could fit the
observations with a $log$ N(He)/N(H) = 1 -- 1.5 model. However, these 
models cannot explain the observed flux deficits in the Spitzer observations. Bergeron et al. (1995) and Hansen (1998)
models cannot explain these flux deficits either (B.M.S. Hansen and P. Bergeron 2005, private communication). 
CIA is expected to create wiggles
and bumps in the spectrum up to 3$\mu$m, and the SEDs of cool white dwarfs are expected to return to normal in the mid-infrared
(Figure 4; see also Figure 6 of Jorgensen et al. 2000). 

Figure 4 also shows a power law distribution with $\alpha=2$, i.e. a Rayleigh-Jeans spectrum (dashed line).
The infrared spectral energy distribution of LHS 1126 follows a Rayleigh--Jeans spectrum
fairly well. It is striking to note that we can fit the ultraviolet and optical part of the SED with a 5400 K blackbody, and the
infrared part of the SED with $T_{\rm eff}>10^5$ K.
The power law is a reasonable but not a perfect fit to the infrared spectral energy
distribution of this object, and photometry in the other IRAC channels or spectroscopy with the Spitzer Telescope
may reveal more structure in the mid-infrared, but the current data suggest the following:

1 -- The CIA opacity calculations are incomplete, and there is a significant, unexplained flux deficit in the mid-infrared.
As the densities in white dwarf atmospheres approach a significant fraction of the liquid state densities, molecular
absorption (H$_2$-H$_2$, H$_2$-He), emission, and light scattering are expected to become increasingly important. 
Current CIA opacity calculations in dense media may be incomplete in certain wavelength regimes, for example
the roto-vibrational band of H$_2$ CIA may be enhanced compared to predictions (L. Frommhold and D. Saumon 2005, private communication).
On the other hand, the featureless spectra of the ultracool white dwarfs (Gates et al. 2004), and the reasonably good
fit of a Rayleigh-Jeans distribution to the LHS 1126 SED in the infrared, suggest another mechanism to explain the mid infrared
flux deficits observed in white dwarfs. 
CIA is expected to disappear  with the dissociation of molecules when $T_{\rm eff}$ increases past 5500 K. Therefore,
white dwarfs with $T_{\rm eff}>6000K$ that show infrared flux deficits could not be explained by changing the CIA opacities.

2 -- The mid-infrared flux deficits are caused by some as-yet unrecognized physical process(es). The problem may be caused by an unknown
or poorly understood absorption process, or  it may be the result of a different source function operating in these dense
atmospheres. H$^-$ bound-free absorption is thought to be the most important source of opacity in warmer DA white dwarfs. More work
is required to test if possible changes in this absorption could help explain the mid-infrared flux deficits.
The input physics used in white dwarf model atmospheres is mostly based on the ideal gas approximation. The extreme conditions in white
dwarf atmospheres require a new look at dense medium effects on the equation of state, chemistry, opacities, and radiative
transfer. Refractive opacities, presence of heavy elements,
or formation of trace species and many other factors can change the opacity sources in white dwarf atmospheres (see Kowalski 2005
for a detailed discussion).

Near/mid-infrared photometry and spectroscopy of ultracool white dwarfs and the so-called C$_2$H stars will be useful to test these ideas.
At this time, we only report and do not understand the observed mid-infrared flux deficits, but this mystery needs to be resolved before
we can use the cool white dwarfs as accurate chronometers to find the age of Galactic populations.

\acknowledgements
We would like to thank Didier Saumon, Lothar Frommhold, Jason Kalirai, Brad Hansen, and Pierre Bergeron for helpful discussions,
and our referee, Jay Farihi, for helpful suggestions that greatly improved the article.
This work is based in part on observations made with the $Spitzer$ $Space$ $Telescope$, which is operated by the Jet Propulsion
Laboratory, California Institute of Technology under NASA contract 1407. Support for this work was provided by NASA through
award Project NBR: 1269551 issued by JPL/Caltech to the University of Texas.
This publication makes use of data products from the Two Micron All Sky Survey, which is a joint project of the University of Massachusetts and the Infrared Processing and Analysis Center/California Institute of Technology, funded by the National Aeronautics and Space Administration and the National Science Foundation.

\clearpage
%TABLE1
\begin{deluxetable}{llrrrrcc}
%\rotate
\tabletypesize{\footnotesize}
\tablecolumns{8}
\tablewidth{0pt}
\tablecaption{Infrared Photometry of Cool White Dwarfs}
\tablehead{
\colhead{Object}&
\colhead{Spectral Type}&
\colhead{$T_{\rm eff}$(K)$^1$}&
\colhead{$F_J$(mJy)}&
\colhead{$F_H$(mJy)}&
\colhead{$F_K$(mJy)}&
\colhead{$F_{4.5\mu}$(mJy)}&
\colhead{$F_{8\mu}$(mJy)}
} \startdata
WD0009+501&DAP & 6540/6683 & 6.40 $\pm$ 0.12 & 5.14 $\pm$ 0.10 & 3.53 $\pm$ 0.07 &  0.94 $\pm$ 0.05 & 0.32 $\pm$ 0.02\\
WD0018$-$267& DA & \nodata/5498 & 15.88 $\pm$ 0.29 & 14.72 $\pm$ 0.30 & 10.51 $\pm$ 0.20 & 2.79 $\pm$ 0.09 & 1.02 $\pm$ 0.04\\
WD0038$-$226& C$_2$H:& 5400/\nodata & 7.34 $\pm$ 0.14 & 4.14 $\pm$ 0.09 & 2.13 $\pm$ 0.05 & 0.48 $\pm$ 0.02 & 0.19 $\pm$ 0.03\\
WD0101+048$^2$ & DA& 8080/8160 & 6.32 $\pm$ 0.12 & 4.49 $\pm$ 0.09 & 2.86 $\pm$ 0.06 & 0.75 $\pm$ 0.03 & 0.29 $\pm$ 0.03\\
WD0126+101$^2$ & DA& 8500/8700 & 3.89 $\pm$ 0.07 & 2.69 $\pm$ 0.06 & 1.73 $\pm$ 0.04 & 0.45 $\pm$ 0.02 & 0.16 $\pm$ 0.07\\
WD0141$-$675&DA & \nodata/6469 & 11.37 $\pm$ 0.20 & 8.85 $\pm$ 0.18 & 6.20 $\pm$ 0.12 & 1.64 $\pm$ 0.06 & 0.59 $\pm$ 0.03\\
WD0552$-$041& DZ & 5060/5016 & 9.63 $\pm$ 0.18 & 7.35 $\pm$ 0.15 & 5.17 $\pm$ 0.10& 1.80 $\pm$ 0.06& 0.72 $\pm$ 0.07\\
WD0553+053&DAP & 5790/5853 & 10.73 $\pm$ 0.19 & 8.36 $\pm$ 0.17 & 5.79 $\pm$ 0.11 & 1.64 $\pm$ 0.05 & 0.51 $\pm$ 0.04\\
WD0752$-$676& DA & 5730/5774 & 12.94 $\pm$ 0.23 & 10.47 $\pm$ 0.21 & 7.57 $\pm$ 0.15 & 2.06 $\pm$ 0.08 & 0.77 $\pm$ 0.04\\
WD0839$-$327$^2$ & DA& \nodata/8978 & 37.26 $\pm$ 0.68 & 24.81 $\pm$ 0.51 & 16.04 $\pm$ 0.32 & 4.01 $\pm$ 0.14 & 1.45 $\pm$ 0.06\\
WD0912+536&DCP & 7160/7273 & 7.57 $\pm$ 0.14 & 5.32 $\pm$ 0.11 & 3.72 $\pm$ 0.07 & 1.10 $\pm$ 0.04 & 0.41 $\pm$ 0.03\\
WD1055$-$072& DC & 7420/7252 & 4.95 $\pm$ 0.09 & 3.45 $\pm$ 0.07 & 2.69 $\pm$ 0.06 & 0.64 $\pm$ 0.02 & 0.25 $\pm$ 0.03\\
WD1121+216$^3$ & DA & 7490/7540 & 5.93 $\pm$ 0.11 & 4.39 $\pm$ 0.09 & 2.91 $\pm$ 0.06 & 0.81 $\pm$ 0.03 & 0.34 $\pm$ \nodata \\
WD1223$-$659& DA & \nodata/7793 & 7.39 $\pm$ 0.14 & 5.10 $\pm$ 0.12 & 3.20 $\pm$ 0.07 & 0.90 $\pm$ 0.04 & 0.36 $\pm$ 0.07\\
WD1748+708& DXP & 5590/5964& 13.15 $\pm$ 0.24 & 9.98 $\pm$ 0.20 & 6.62 $\pm$ 0.13 & 1.89 $\pm$ 0.06 & 0.70 $\pm$ 0.04\\
WD1756+827& DA & 7270/7285 & 5.61 $\pm$ 0.10 & 4.18 $\pm$ 0.09 & 2.83 $\pm$ 0.06 & 0.75 $\pm$ 0.02 & 0.26 $\pm$ 0.04\\
WD1953$-$011& DAP & 7920/7851 & 9.43 $\pm$ 0.17 & 6.29 $\pm$ 0.13 & 4.15 $\pm$ 0.09 & 1.13 $\pm$ 0.04 & 0.41 $\pm$ 0.02\\
WD2140+207 & DQ& 8860/9407 & 10.24 $\pm$ 0.18 & 6.90 $\pm$ 0.14 & 4.52 $\pm$ 0.09 & 1.19 $\pm$ 0.05 & 0.41 $\pm$ 0.03\\
\enddata
\tablecomments{(1) Estimated $T_{\rm eff}$ from BLR vs. this study. (2) Suspected or known double degenerate. (3) 8 $\mu$m photometry of WD1121+216 is affected by a nearby star.}
\end{deluxetable}

\clearpage
\begin{figure}
\includegraphics[angle=-90,scale=.7]{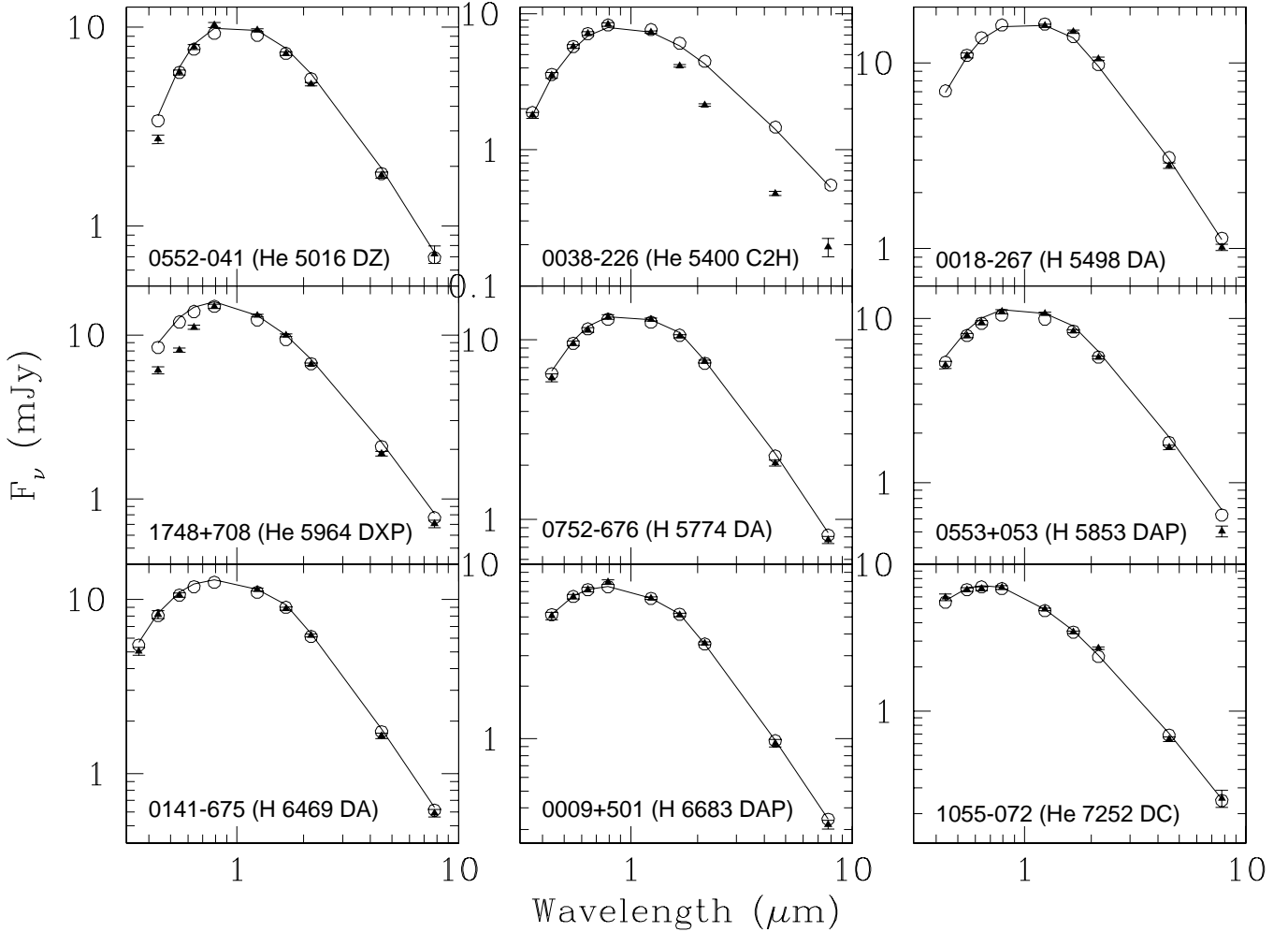}
\caption{Spectral energy distributions of cool white dwarfs observed with the Spitzer Space Telescope. The observed
fluxes are shown as filled triangles with errorbars, whereas the expected flux distributions from synthetic photometry
of white dwarf models (kindly made available to us by D. Saumon and D. Koester)
are shown as open circles.
Solid lines represent the same models normalized to the observed SEDs in the $J$-band
The object name, atmospheric composition, $T_{\rm eff}$, and spectral type are given in each panel.}
\end{figure}

\clearpage
\begin{figure}
\figurenum{1}
\includegraphics[angle=-90,scale=.7]{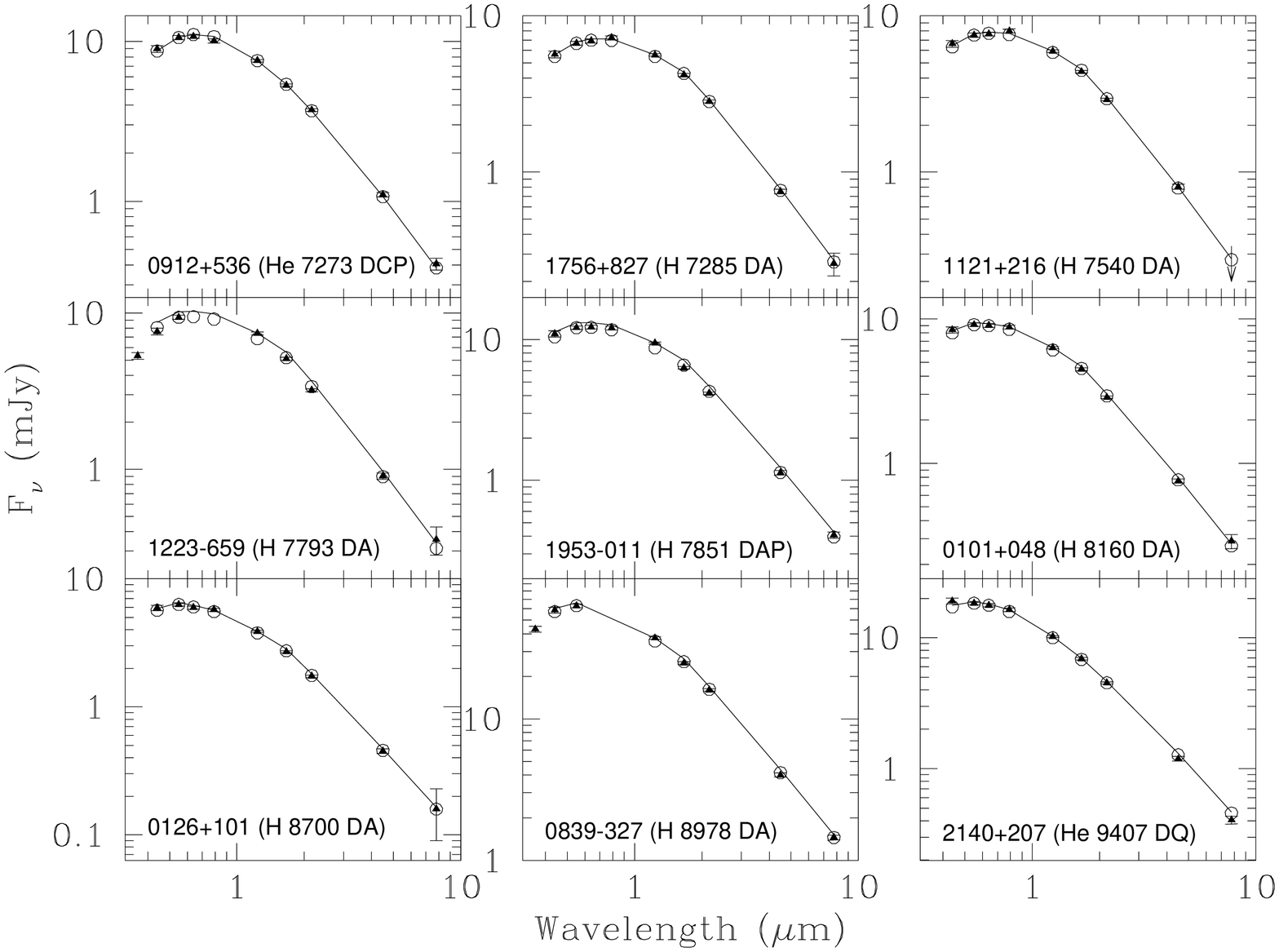}
\caption{contd.}
\end{figure}

\clearpage
\begin{figure}
\includegraphics[angle=0,scale=.8]{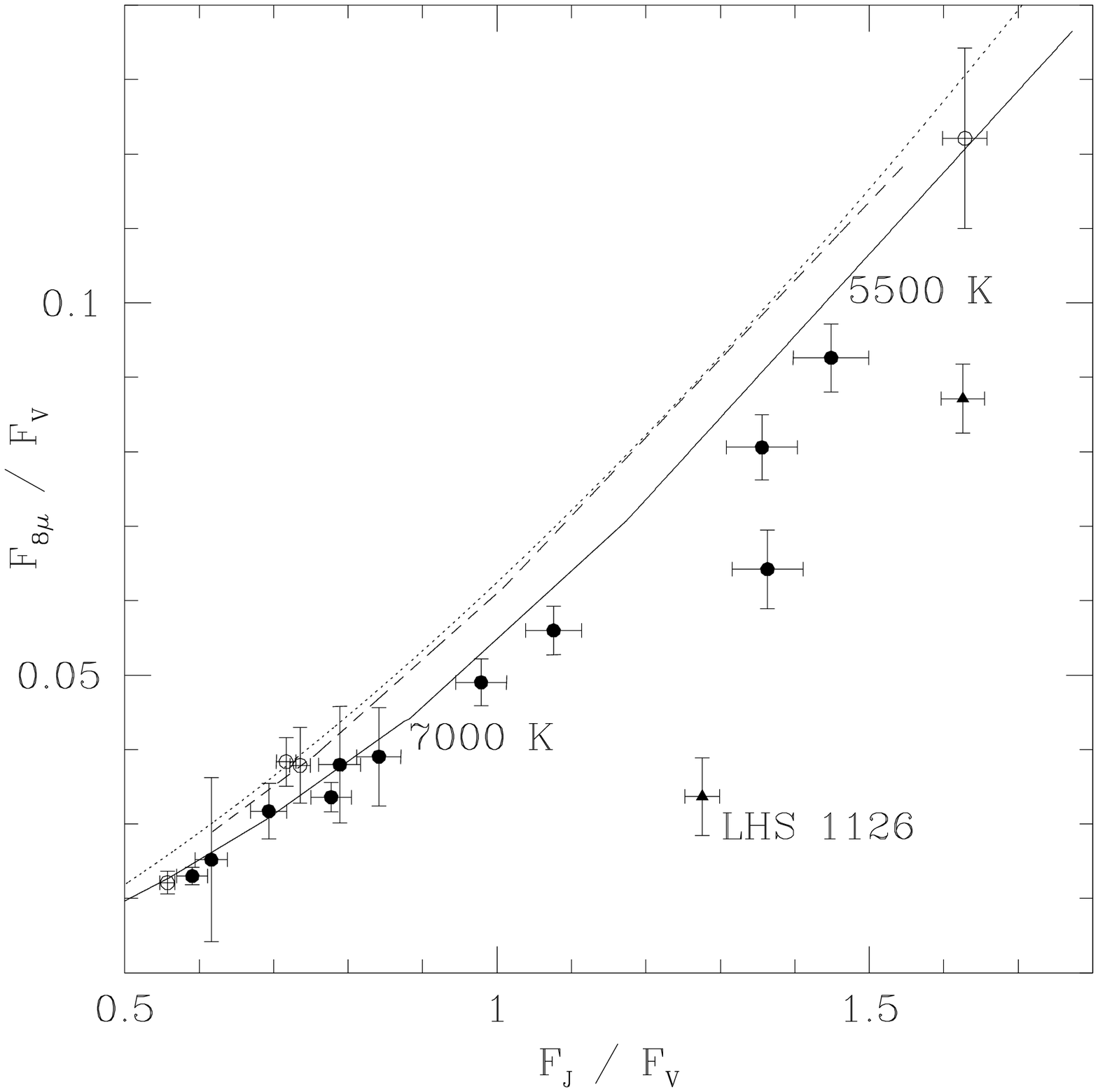}
\caption{Ratio of observed fluxes in $V$, $J$, and 8$\mu$m bands for DA (filled circles) and DB (open circles) white dwarfs.
White dwarfs with uncertain spectral types are shown as filled triangles.
Expected cooling tracks for DA (5000 -- 60000 K, solid line) and
DB white dwarfs (5000 -- 8000 K, dashed line) are also shown. A dotted line shows the expected sequence for a blackbody
(5000 -- 10$^5$ K).}
\end{figure}

\clearpage
\begin{figure}
\includegraphics[angle=0,scale=.8]{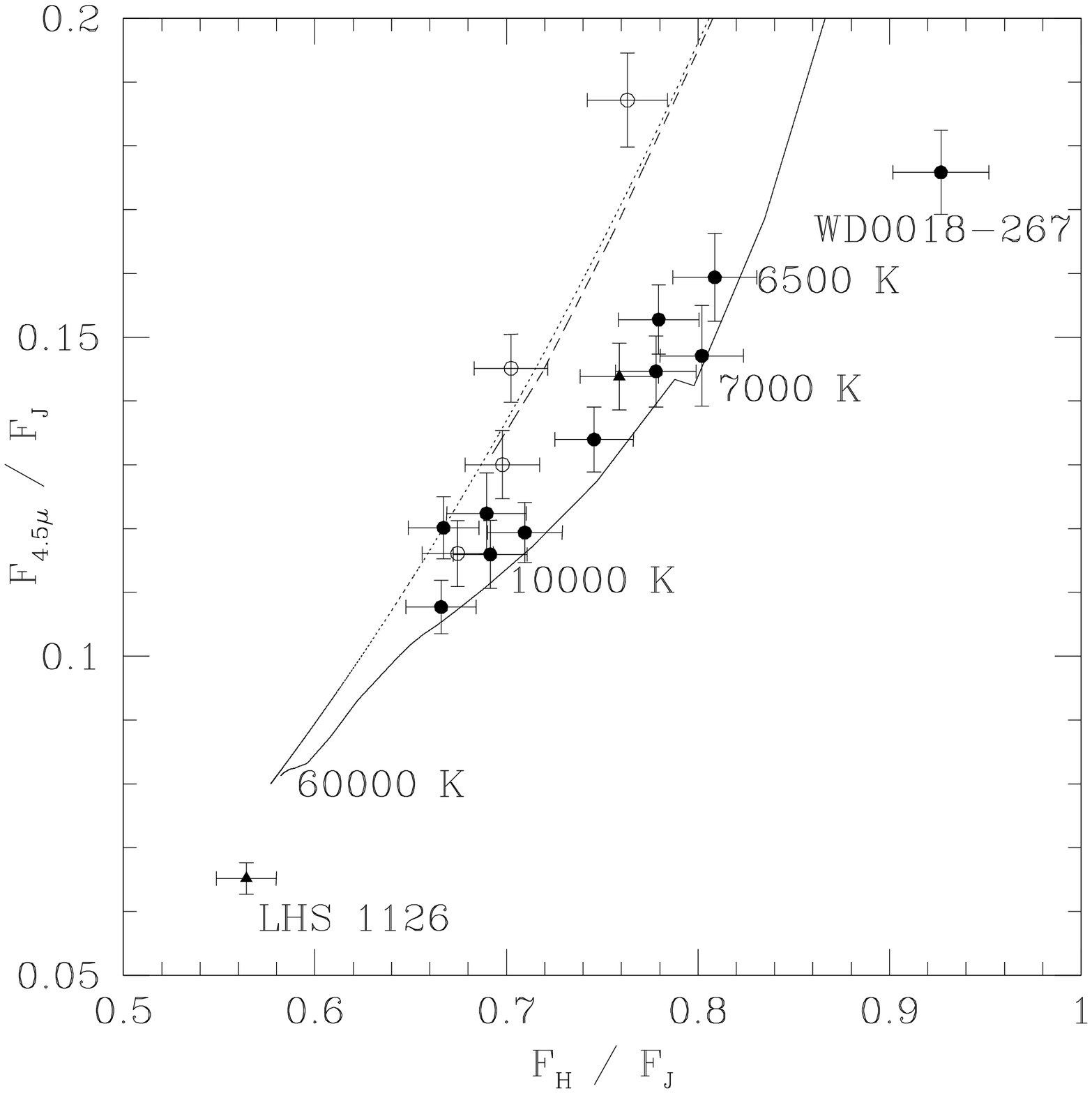}
\caption{Ratio of observed fluxes in $J$, $H$, and 4.5$\mu$m bands. The symbols are the same as in Figure 2.
Note that the discontinuity at 7000 K is due to transition from Koester models to Saumon models.}
\end{figure}

\clearpage
\begin{figure}
\includegraphics[angle=-90,scale=.7]{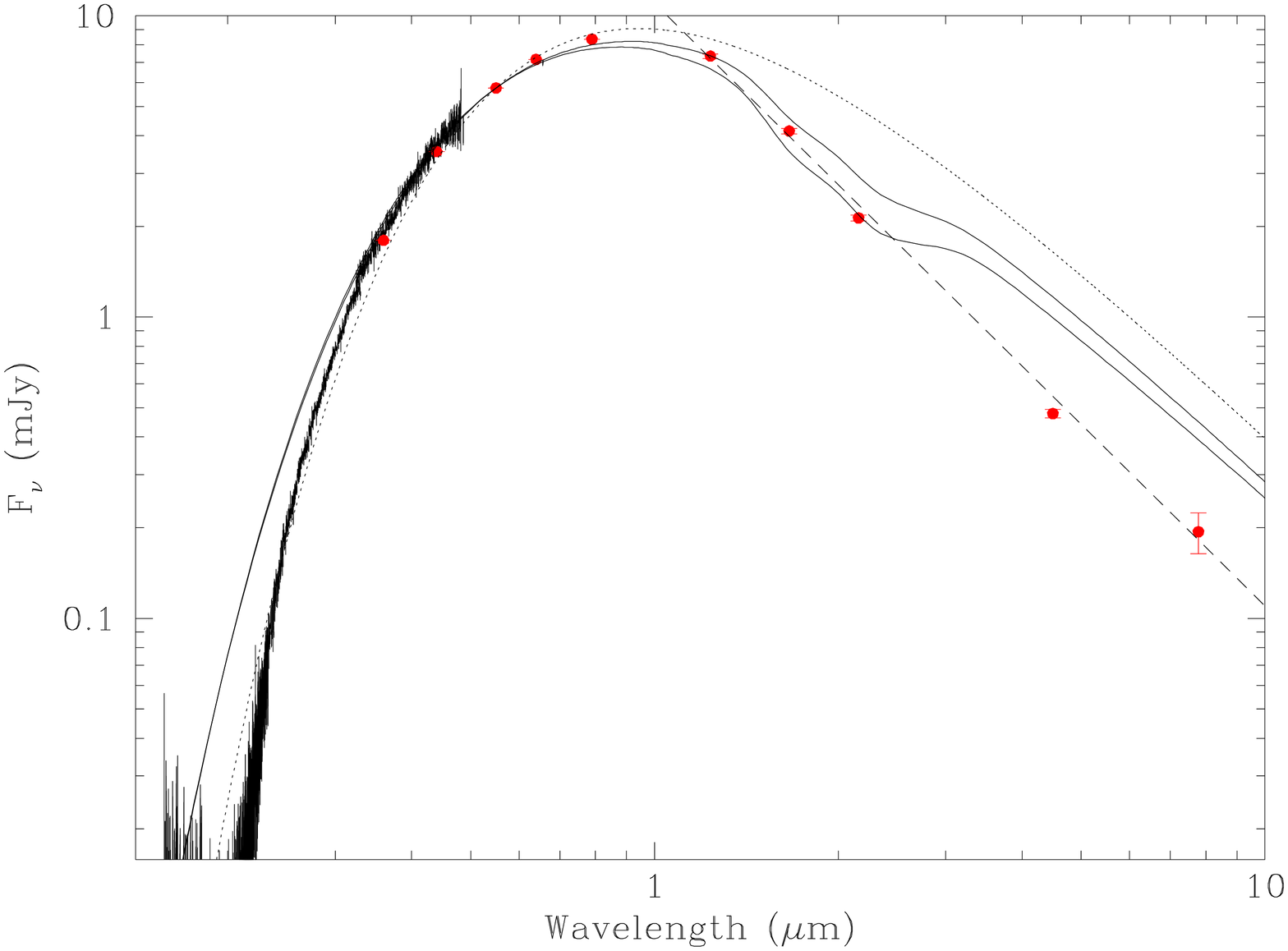}
\caption{Spectral energy distribution of LHS 1126, along with a 5400 K blackbody (dotted line).
White dwarf models with $T_{\rm eff}=$5400 K, $log$ g=7.9, and $log$ N(He)/N(H) = 1.5 (lower solid line) and $log$ N(He)/N(H)
= 1 (upper solid line) are also shown. The dashed line shows a power law with $\alpha=2$.}
\end{figure}

\end{document}